# Vapor-phase growth and characterization of van der Waals $Bi_{2-x}Sb_xTe_{3-y}Se_y$ platelets on semiconducting $MoS_2$

M. Zhezhu[1*], V. Torosyan[1], V. Mehrabyan[1], A. Vasil'ev[1]

[1]*A.B. Nalbandyan Institute of Chemical Physics, Yerevan, Republic of Armenia*

**Correspondence should be addressed to* marina.zhezhu@ichph.sci.am

**Abstract**

Three-dimensional topological insulators of the tetradymite family, such as $Bi_{2-x}Sb_xTe_{3-y}Se_y$ (BSTS), are attractive for spintronic and quantum device applications because of their compensated, bulk-insulating character. Nanostructured BSTS platelets are promising for practical TI-based devices. However, their growth often requires complex equipment, organometallic precursors, or electrically insulating substrates. Simple growth on technologically relevant semiconducting substrates remains largely unexplored. Here, we report the vapor-phase growth of van der Waals BSTS platelets directly on semiconducting $MoS_2$ using a pre-synthesized $BiSbTe_{1.5}Se_{1.5}$ crystal as the source material. The grown BSTS platelets exhibit thicknesses ranging from 4 nm to 246 nm, with an average lateral size of 1.4 $\mu$m for individual platelets. Raman spectra of platelets, acquired in correlation with AFM thickness measurements, reveal contributions from both BSTS and the underlying $MoS_2$ substrate. Among the characteristic BSTS modes, $A^2_{1g}$ mode shows the highest sensitivity to platelet thickness and local composition. EDS analysis reveals a systematic thickness dependence of the platelet composition, with an apparent crossover around 49 nm that coincides with a change in the thickness dependence of the low-wavenumber component of the $E^2_g$ Raman band. This correlation suggests a composition-related origin of the Raman behavior, with the compositional variations tentatively attributed to the differential volatility of the constituent elements during growth. These results demonstrate that $MoS_2$ is a promising van der Waals platform for BSTS deposition and provide a structural and compositional baseline for future studies of BSTS/$MoS_2$ heterostructures.

## 1. Introduction

Three-dimensional topological insulators (3D TIs) are quantum materials in which strong spin-orbit coupling induces an inverted bulk band structure, giving rise to spin-momentum locked, gapless surface states that reside within the bulk band gap [1]. Among the various classes of 3D TIs, tetradymite compounds are particularly research interest as a versatile platform for exploring topological surface states and their potential applications in spintronics [2, 3], quantum information processing [4], thermoelectric energy conversion [5–7], and other emerging device concepts [8], although the tellurium content of these compounds has also motivated parallel research into Te-free alternatives [9, 10]. Recent studies have focused on quaternary tetradymite compounds $Bi_{2-x}Sb_xTe_{3-y}Se_y$ (BSTS) [3, 5, 11-16], because the prototypical binary chalcogenides ($Bi_2Se_3$, $Bi_2Te_3$, and $Sb_2Te_3$) commonly suffer from unintentional bulk doping, arising from native point defects, which makes it difficult to isolate the topological surface-state transport from parasitic bulk conduction and thereby limits their implementation in electronic devices [17-19]. In BSTS, compensation between donor-and acceptor-type defects can suppress bulk conduction and enhance the relative contribution of the topological surface states. However, the effectiveness of this compensation is sensitive to stoichiometry and atomic-site disorder [13, 20]. Establishing how the local composition develops during growth is therefore particularly important for understanding the structural and electronic properties of BSTS nanostructures.

A promising route toward the practical realization of TI-based devices is the growth of nanostructured plates, characterized by micrometer-scale lateral dimensions combined with nanometer-scale thickness, which are attractive both for probing topological surface properties and for the fabrication of novel devices [21-23]. Molecular beam epitaxy (MBE), commonly used for BSTS growth, provides superior control over thickness and stoichiometry [2, 24, 25] but requires expensive, technically demanding equipment. Alternative approaches to BSTS growth remain limited and include metal-organic vapor-phase epitaxy (MOVPE), which relies on toxic and often unstable metal-organic precursors [26], and vapor solid and chemical vapor transport methods, which have predominantly employed electrically insulating substrates for plate growth [21, 22]. Among these, fluorophlogopite mica has proven particularly effective in growing laterally extended, thin nanoscale plates [21]. However, its electrically insulating nature limits direct access to an electronically active substrate-platelet interface and may necessitate transferring the grown material for certain device configurations. In addition, natural mica exhibits compositional variability [27], which can compromise its optical uniformity and complicate reliable optical and photonic characterization of the grown material. Layered metal dichalcogenides, in contrast, combine a dangling-bond-free surface favorable for van der Waals epitaxy with semiconducting functionality [28]. They therefore represent a promising class of substrates for direct growth of BSTS platelets and the subsequent investigation of interfacial charge transfer, band alignment, and carrier transport. Nevertheless, to the best of our knowledge, the direct vapor-phase growth of quaternary BSTS platelets on such substrates has not previously been reported.

In this work, we report the vapor-phase growth of $Bi_{2-x}Sb_xTe_{3-y}Se_y$ platelets directly on a semiconducting $MoS_2$ substrate using a pre-synthesized $BiSbTe_{1.5}Se_{1.5}$ crystal as the source material. The morphology, thickness, vibrational properties, and local elemental composition of the resulting platelets are systematically characterized by atomic force microscopy (AFM), scanning electron microscopy (SEM) with energy-dispersive X-ray spectroscopy (EDS), and Raman spectroscopy, with particular focus on their compositional variation as a function of platelet thickness, as well as vibrational variation as a function of both platelet thickness and the local Bi/Sb and Te/Se atomic ratios.

## 2. Experimental

Prior to vapor phase growth, a bulk $BiSbTe_{1.5}Se_{1.5}$ source crystal was synthesized. The source composition, corresponding to $x = 1.0$ and $y = 1.5$, was deliberately selected as a reference BSTS composition. Its equal Bi/Sb and Te/Se atomic ratios provide a convenient baseline for assessing possible compositional deviations in the deposited platelets. For this purpose, high-purity elemental Bi, Sb, Te, and Se (99.9999%) were mixed in a stoichiometric ratio of Bi:Sb:Se:Te = 1:1:1.5:1.5. The elemental mixture was loaded into a quartz ampoule with an inner diameter of 15 mm, evacuated to 5 Pa, and flame-sealed. The ampoule was heated to 950 °C at 2 °C/min, held at this temperature for 4 h to homogenize the melt, then cooled to 600 °C at 50 °C/min, and finally furnace-cooled to room temperature. The phase purity and crystal structure of the $BiSbTe_{1.5}Se_{1.5}$ crystal were examined by X-ray diffraction (XRD, Rigaku MiniFlex), while its elemental composition was determined by EDS using SEM (Thermo Fisher Scientific).

$MoS_2$ substrates were prepared by mechanically cleaving a bulk $MoS_2$ crystal (2D Semiconductors, USA) mounted on a glass substrate using Nitto 224 SPV adhesive tape (Nitto Denko Corporation, Japan), thereby exposing a fresh, atomically clean van der Waals surface suitable for growth.

BSTS platelets were subsequently grown on freshly cleaved $MoS_2$ substrates in a single-zone tube furnace under continuous pumping at 5 Pa, without a carrier gas, thereby enabling a simple vacuum-assisted growth process. A pre-synthesized $BiSbTe_{1.5}Se_{1.5}$ crystal, ground into powder, was used as the source material. Approximately 100 mg of the source material was placed in a quartz boat positioned in the heating zone. The source-to-substrate distance was fixed at 10 cm, with the $MoS_2$ on a glass substrate positioned at the end of the heating zone inside the furnace tube. The heating zone was ramped to 550 °C over 30 min and maintained at this temperature for 1 h. After the 1 h growth process, the sample was immediately removed from the heating zone to cool to room temperature. A schematic illustration of the experimental process is shown in Figure 1. Preliminary deposition experiments were conducted over a source-temperature range of 500–600 °C, using sapphire, quartz, and mica as test substrates. Representative BSTS platelets obtained during these optimization experiments are shown in Figure S1.

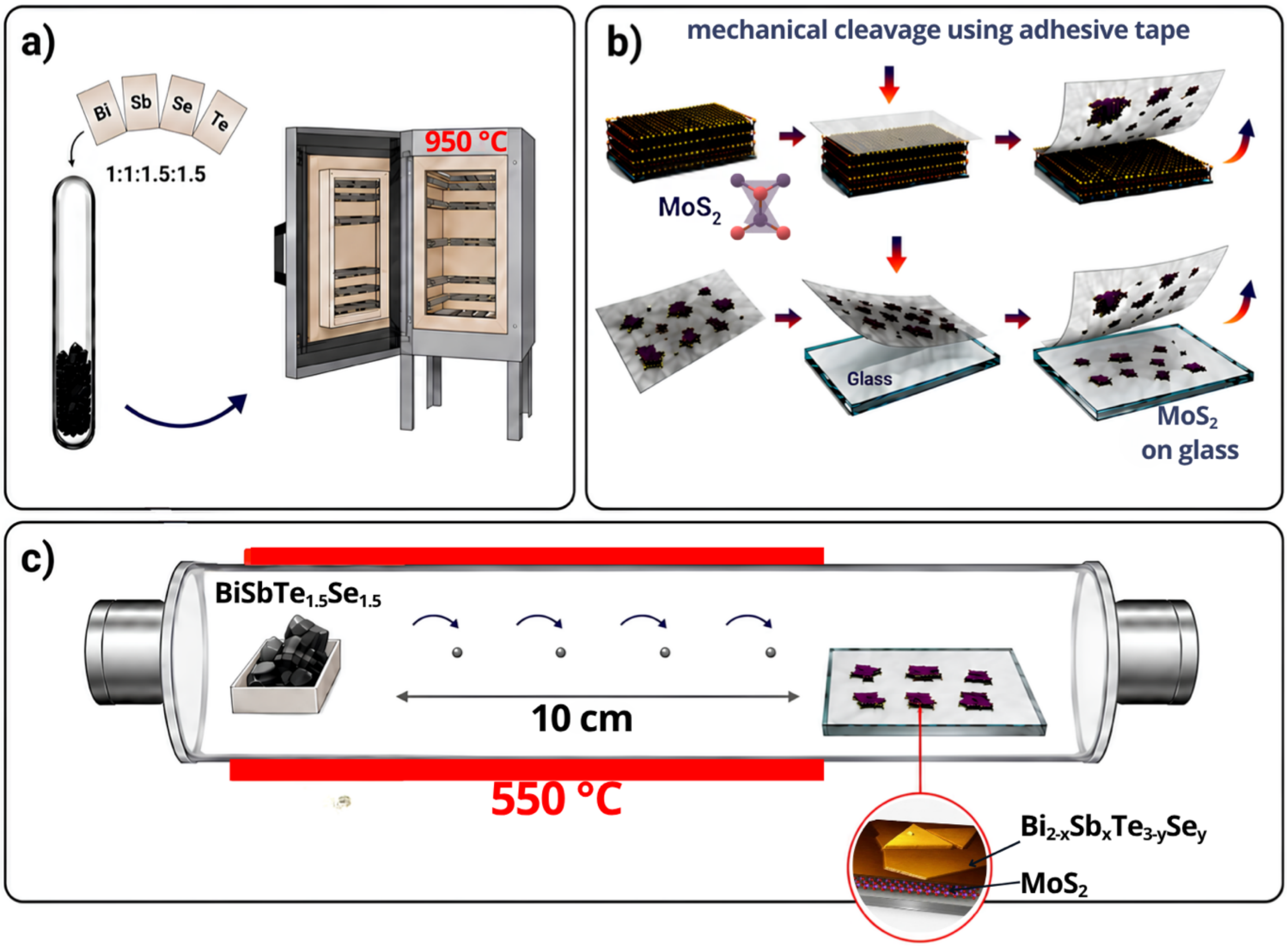

Figure 1. Schematic illustration of BSTS nanoplatelet preparation process: (a) synthesis of the BSTS source crystal in a quartz ampoule, (b) preparation of $MoS_2$ substrates by mechanical cleavage of bulk crystal, and (3) vapor growth of BSTS nanoplatelets on the cleaved $MoS_2$ substrates.

The surface morphology of the samples was characterized by SEM, while correlative AFM and Raman spectroscopy measurements were performed using an AFM-combined Raman system (LabRAM HR Evolution, Horiba) equipped with a Peltier-cooled CCD detector and 633 nm He–Ne laser. AFM was used to determine the topography and thickness of individual platelets. Raman spectra were acquired using a 100× objective, a 600 gr/mm diffraction grating, and a laser power of 0.5 mW. The 633 nm edge and bandpass filter configuration provided access to Raman shifts above 50 $cm^{-1}$. The local elemental compositions of BSTS platelets were determined by SEM–EDS and correlated with their AFM-measured thicknesses. The EDS measurements were performed at an accelerating voltage of 20 kV, and an acquisition time of 30 s. According to the manufacturer, the minimum detection limit of EDS is below 0.1 wt.%.

**3. Results and Discussion**

The preliminary ground $BiSbTe_{1.5}Se_{1.5}$ single crystal exhibited phase purity and can be indexed to $R\bar{3}m$ space group with lattice parameters of $a$=$b$=4.1993 Å and $c$=29.73 Å (Figure S2(a)). EDS elemental analysis confirmed that the crystal composition is in good agreement with the nominal Bi:Sb:Te:Se ratio of 1:1:1.5:1.5 (Figure S2(b)). Raman analysis in Figure S2(c) revealed that its vibrational modes are located between those of the parent compounds $Bi_2Te_3$ and $Sb_2Te_3$, with peak positions closer to those of $Sb_2Te_3$ [29-31]. The Raman spectrum of the $BiSbTe_{1.5}Se_{1.5}$ crystal exhibited the out-of-plane $A^1_{1g}$ and $A^2_{1g}$ modes at 65.3 and 164.3 $cm^{-1}$, respectively, together with the in-plane $E^2_g$ mode at 119.4 $cm^{-1}$. The low-frequency $E^1_g$ mode, typically expected below 50 $cm^{-1}$, is usually very weak and was not observed due to the low-frequency cutoff of the optical filter configuration. A disordered solid solution with randomly distributed Bi/Sb and Te/Se atoms could potentially activate additional Raman modes or induce splitting of the characteristic bands [30]. Nevertheless, the Raman spectrum of $BiSbTe_{1.5}Se_{1.5}$ crystal exhibited three dominant vibrational features, with compositional disorder primarily manifested as peak broadening (FWHM of $E^2_g$ mode is 35.6). Earlier experimental Raman studies also showed three peaks for BSTS compositions [21, 22, 32]. Raman analysis of the cleaved $MoS_2$ substrate in Figure S2(d) revealed characteristic vibrational modes, with $E^1_{2g}$ and $A_{1g}$ peaks located at 379.1 $cm^{-1}$ and 405.7 $cm^{-1}$, respectively [33]. The Raman shift difference between these modes ($\Delta$=26.6 $cm^{-1}$) corresponds to the bulk $MoS_2$ [34], indicating the presence of multilayer flakes. The enlarged region from 100 to 200 $cm^{-1}$ indicated that vibrational modes within this range may potentially contribute to the Raman response of the grown BSTS platelets on $MoS_2$ substrate. This region contains characteristic bands at 124.4, 136.4, 151.3, and 177.8 cm$^{-1}$, which can be assigned to the $A_{1g}(\Gamma)$-$E_{1g}(\Gamma)$, $E_{1g}(M)$-$ZA(M)$, $TA(M)$, $A_{1g}(M)$-$LA(M)$, respectively [33].

The initial evaluation of vapor-grown van der Waals BSTS on a semiconducting $MoS_2$ substrate by optical microscopy revealed a pronounced optical contrast between the deposited material and the $MoS_2$ surface (Figure 2(a)).

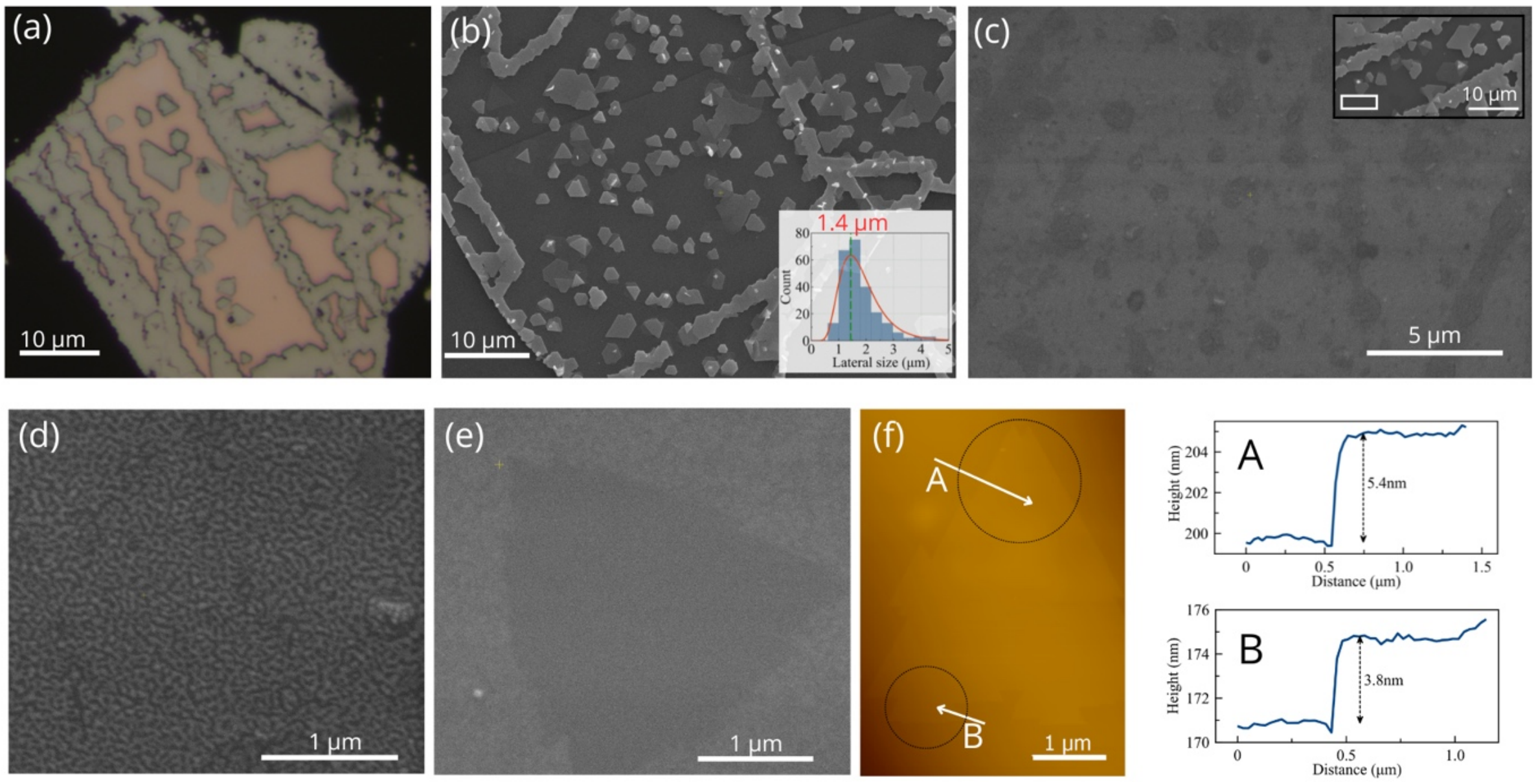

Figure 2. Microscopic characterization of vapor-phase-grown van der Waals BSTS on a semiconducting $MoS_2$ substrate: (a) optical microscopy image, (b) SEM image with the corresponding histogram of the lateral size distribution of the BSTS platelets, (c) SEM image of the initial nucleation region, corresponding to the boxed are shown in the inset, (d) enlarged view of the nucleation area, and characterization of individual triangular platelets by (e) SEM and (f) AFM, with the corresponding thickness estimation.

Examination of different regions of the same sample revealed that the plates exhibit hexagonal or truncated-trigonal morphologies typical of their compound family [31, 35, 36]. In addition to regions containing platelets with an average lateral size of approximately 1.4 $\mu$m (see Figure 2(b)), other areas exhibit pronounced morphological heterogeneity, comprising isolated platelets, stacked platelets, and densely packed clusters (see Figure 2(a)). SEM and AFM analyses of the $MoS_2$ surface in this region (Figure 2(c, d)) reveal the initial stages of nucleation, along with isolated triangular platelets with thicknesses of approximately 4–5 nm. These observations suggest that BSTS nucleation proceeds *via* the formation of ultrathin, laterally confined islands, which subsequently evolve into thicker, more extended structures. Such behavior can be rationalized by considering the surface properties of layered $MoS_2$ and the anisotropic crystal structure of tetradymite-type BSTS. The basal plane of $MoS_2$ is relatively free of dangling bonds and therefore provides a weakly interacting surface for the adsorption and migration of vapor-phase species [28]. At the same time, local surface imperfections, including chalcogen vacancies, step edges, and other defects, may provide energetically favorable sites for the trapping of adsorbed species and the formation of stable nuclei [37]. $MoS_2$ surface may provide favorable conditions for mobile diffusion of adsorbed species, facilitating their transport toward stable nuclei and subsequent lateral growth [38]. Once stable BSTS nuclei are formed, the strongly anisotropic layered structure of tetradymite-type materials favors lateral growth along the basal plane [39], consistent with the characteristic platelet morphology. The 4–5 nm triangular platelets observed in Figure 2(e, f) can therefore be associated with an early island-growth stage, followed by lateral expansion and subsequent thickening with continued material supply.

For detailed AFM–Raman characterization, a region exhibiting pronounced growth heterogeneity was intentionally selected. This region, shown in Figure 3, exhibits a wide thickness distribution (25–246 nm) within a confined lateral area. Such a region was deliberately selected to enable a systematic, statistically robust comparison of Raman spectral characteristics across a broad range of platelet thicknesses obtained under nearly identical local growth conditions. This approach reduces the influence of spatially varying growth conditions (e.g., temperature and precursor concentration gradients) that are inherently present when comparing regions located at different positions on the sample. However, this raises questions about the origin of the observed thickness variation. Nucleation may occur at different times and with different local densities across the cleaved $MoS_2$ surface, resulting in unequal growth durations and differences in the amount of material collected by individual islands. Local variations in structure and defect density may further influence nucleation and surface diffusion, thereby contributing to the broad platelet thickness distribution. Studies of the related $Bi_2Te_3/WS_2$ system [37] have demonstrated the sensitivity of platelet nucleation to substrate defects, with an increased nucleation density of few-layer $Bi_2Te_3$ sheets observed on defective $WS_2$ surfaces. Notably, a similarly broad thickness range (~18–250 nm) has recently been observed for $Bi_2Te_3$ crystals grown on $MoS_2$ by chemical vapor transport, where the crystal thickness was found to depend strongly on the growth conditions and the evolution toward thicker crystals was described as consistent with a Volmer–Weber-type growth mode [38]. Different growth mechanisms of $Bi_2Te_3$-based nanostructures on $MoS_2$, including Volmer-Weber-type growth, have also been discussed previously [40]. In this growth mode, adatoms preferentially bond to each other, leading to island nucleation and subsequently crystal growth. This mechanism may also contribute to the broad thickness distribution observed in the present BSTS platelets.

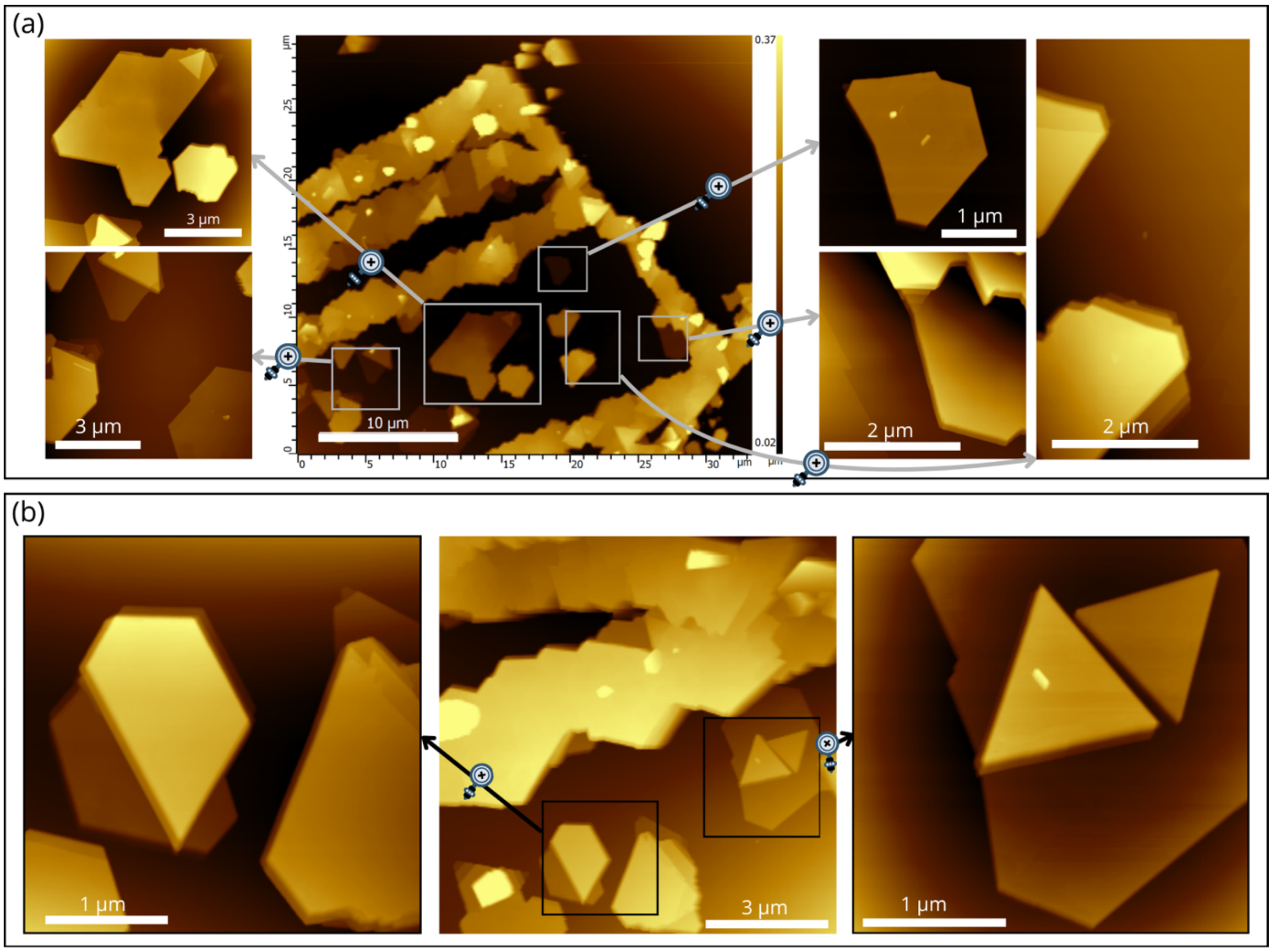

Figure 3. AFM topography maps of the investigated area of the van der Waals growth of BSTS on $MoS_2$ substrate: (a) the main region of the investigated area; (b) the adjacent extension of the same region. Enlarged views highlight representative individual and vertically stacked nanosheets.

In total, 20 plates presented in Figure 3 were investigated by Raman spectroscopy, and the corresponding spectra are shown in Figure S3. Each spectrum was deconvoluted using Gauss fitting. Spectra corresponding to thin, intermediate, and thick platelets within the studied area, as well as reference Raman spectra of the source $BiSbTe_{1.5}Se_{1.5}$ crystal and of mechanically cleaved $MoS_2$, are shown in Figure 4.

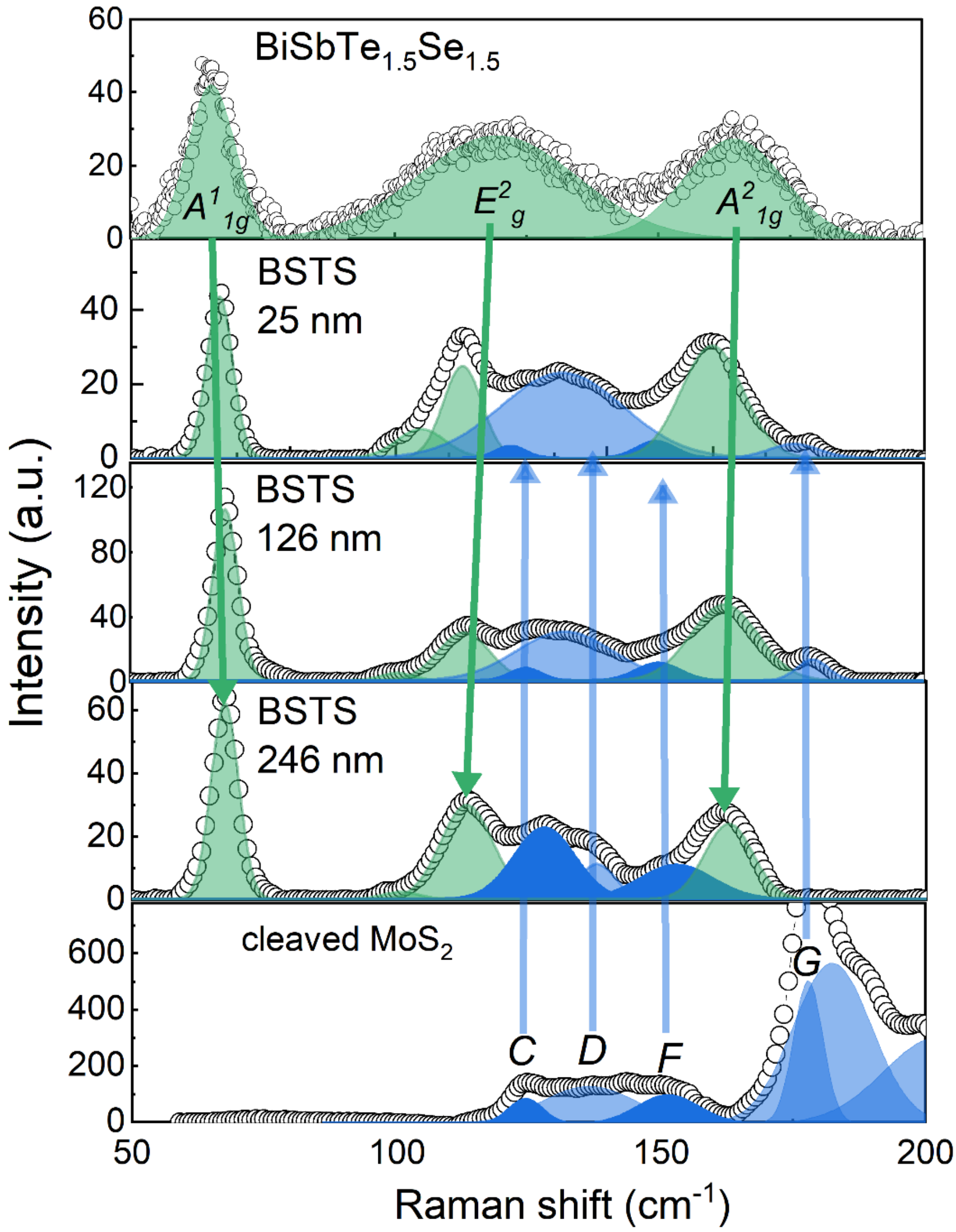

Figure 4. Deconvoluted Raman spectra of the $BiSbTe_{1.5}Se_{1.5}$ crystal, BSTS platelets grown on $MoS_2$ with representative thicknesses of 25 nm (thin), 126 nm (intermediate), and 246 nm (thick), together with the cleaved $MoS_2$ substrate before BSTS deposition. Green and blue peaks correspond to the Raman modes of BSTS and $MoS_2$, respectively. Arrows indicate the evolution of the corresponding Raman bands.

The deconvoluted Raman spectra of grown BSTS on a $MoS_2$ substrate exhibit seven Raman-active phonon modes, whose assignments and positions are summarized in Table 1. Three main modes, $E^2{}_g$, $A^1{}_{1g}$, and $A^2{}_{1g}$ are characteristic modes of tetradymite compounds [30, 41-43] and are also observed in the spectrum of the source material. Four additional peaks, labeled as *C*, *D*, *F*, and *G*, are attributed to the $MoS_2$ substrate discussed above. Notably, the broad and asymmetric line shape of $E^2{}_g$ band can be represented by two components. Similar two-component behavior of $E_g$ -mode has previously been reported in mixed-anion tetradymite compounds and attributed to the coexistence of Se- and Te-terminated local environments within the compositionally disordered anion sublattice [43]. Additionally, DFT calculations showed that BSTS-type compounds can exhibit a large number of additional vibrational modes associated with local inversion-symmetry breaking caused by the random distribution of Bi/Sb and Te/Se atoms over crystallographically inequivalent sites within the quintuple layer [30].

The positions of the three characteristic Raman modes observed for the BSTS plates, $A^1_{1g}$, $E^2_g$, and $A^2_{1g}$ were found to be in good agreement with previously reported values [23, 24, 30, 43]. Relative to the $BiSbTe_{1.5}Se_{1.5}$ source crystal, $A^1_{1g}$ mode of BSTS platelets is shifted toward higher wavenumbers, whereas $E^2_g$ and $A^2_{1g}$ modes are shifted toward lower wavenumbers (see Figure 4), suggesting changes in the local bonding environment. Raman shifts reported for BSTS and $Bi_2Te_3$ have been associated with variations in lattice parameters, composition, and interlayer interactions, suggesting that the observed mode-dependent shifts may arise from a combination of modified van der Waals coupling, local bonding environments, and structural relaxation associated with the platelet geometry [24, 44].

Overall, the Raman spectra of grown BSTS on the $MoS_2$ substrate represent a superposition of vibrational modes originating from both the deposited BSTS platelets and the underlying $MoS_2$ substrate, indicating that the substrate signal remains optically accessible even through plates as thick as 246 nm.

Notably, the phonon position of the $A^2_{1g}$ mode exhibits greater variation (see Table 1). This mode corresponds to a symmetric out-of-plane stretching vibration and, owing to its displacement component along the crystallographic *c*-axis, may be sensitive to the thickness and interfacial environment of the BSTS platelets. Thickness- and interface-related effects on Raman phonons have previously been reported for $Bi_2Se_3$ nanoflakes, where the Raman response of $A_{1g}$ phonons was found to depend on the flake thickness over the studied range of 7-13 QL (~7-13 nm) and on interfacial deformation transfer [24]. In addition, reports on binary $Bi_2Se_3$ and $Bi_2Te_3$ nanoplates showed that pronounced thickness-dependent Raman shifts attributed to phonon confinement are typically observed only below ~15–40 nm [31, 45]. Although the BSTS platelets investigated in the present study are considerably thicker, the thickness-dependent Raman analysis shown in Figure 5(a) reveals a slight increase in the $A^2_{1g}$ mode position with increasing platelet thickness, whereas the positions of the $A^1_{1g}$ and $E^2_g$ characteristic BSTS modes remain nearly unchanged.

In the present data, the low-wavenumber component of the $E^2_g$ band decreases over the 25–49 nm thickness range, while showing a slight tendency to increase over the 49–246 nm range.

Table 1. Raman vibrational modes and their assignments in the 50-200 $cm^{-1}$ range for van der Waals grown BSTS on a $MoS_2$ substrate.

| Compound | Peak label | Raman shift range, $cm^{-1}$ | Mode assignment |
|---|---|---|---|
| BSTS | $A^1_{1g}$ | 66-68 | low frequency out-of-plane mode, arising from movement in phase of the outer layers relative to the central chalcogen atom [46] |
| | $E^2_g$ | 97-105 & 111-114 | a doubly degenerate, high-frequency mode representing in-plane bending vibration (along the *ab* plane), arising from movement in opposite directions of the pnictogen and outer chalcogen atoms [29, 46] |
| | $A^2_{1g}$ | 155-166 | high-frequency, symmetric out-of-plane stretching mode (along the plane perpendicular to *c* axis), arising from movement in opposite phase of the pnictogen atom and the outer chalcogen atom [29, 46] |
| $MoS_2$ | *C* | 118-129 | $A_{1g}(\Gamma)$-$E_{1g}(\Gamma)$ [33] |
| | *D* | 129-138 | $E_{1g}(M)$-$ZA(M)$ [33] |
| | *F* | 134-160 | $TA(M)$ [33] |
| | *G* | 171-180 | $A_{1g}(M)$-$LA(M)$ [33] |

Regarding the Raman modes attributed to the $MoS_2$ substrate, their phonon frequencies vary across the investigated spectral range but show no systematic correlation with BSTS plate thickness (Figure 5(b)). The peak labeled *G* is absent in several spectra and, where observed, shows no correlation with thickness.

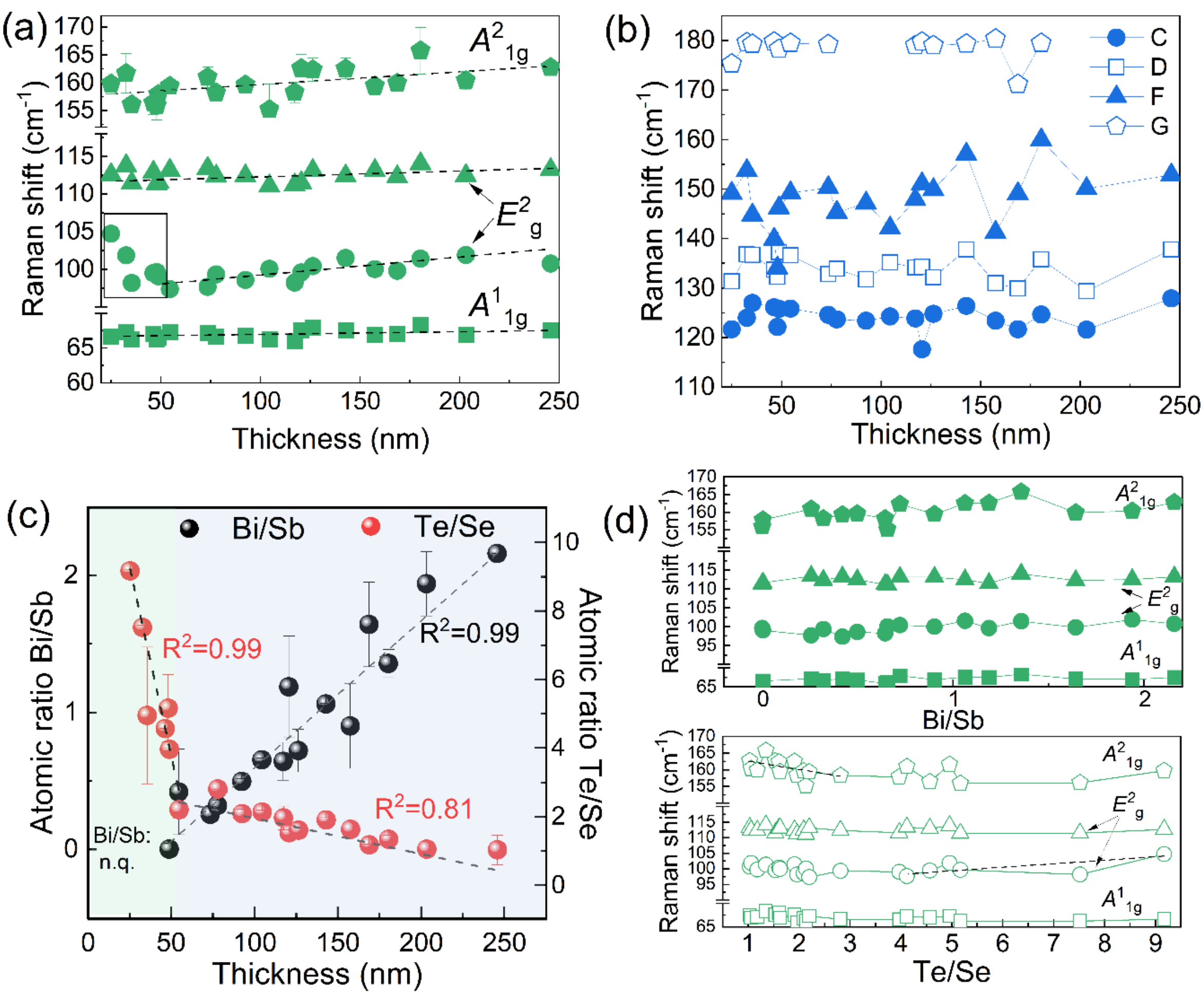

Figure 5. BSTS platelet Raman shift variation with thickness for (a) the characteristic BSTS modes and (b) the characteristic Raman modes of the $MoS_2$ substrate in the studied spectral range of 50–200 $cm^{-1}$. (c) Correlation of the Bi/Sb and Te/Se atomic ratios with BSTS platelet thickness. “Bi/Sb: n.q.” denotes that the Bi/Sb ratio for platelets below 49 nm was not reliably quantifiable under the present EDS measurement conditions. (d) Variation of the Raman shifts of the characteristic BSTS modes with the elemental composition of the platelets. Filled symbols represent the Bi/Sb ratio, while open symbols represent the Te/Se ratio.

To further investigate van der Waals-grown BSTS platelets on $MoS_2$, local EDS analysis was performed on platelets with known thicknesses. The resulting atomic compositions are listed in Table S1. Results of Bi/Sb and Te/Se atomic ratios correlation with BSTS platelet thicknesses are shown in Figure 5(c). It should be noted that quantitative EDS analysis of the thinnest platelets is subject to increased uncertainty. Although Sb, Te, and Se were consistently detected in platelets in the thickness range of 25- 48 nm, Bi was not reliably quantified and therefore is designated as “n.q.” in Table S1. Consequently, the Bi/Sb ratio cannot be reliably determined for these platelets, and the corresponding data are not included in the quantitative Bi/Sb analysis. Despite this limitation, EDS data reveal an apparent change in compositional behavior around 49 nm. Above this thickness, the Bi/Sb ratio increases approximately linearly with platelet thickness, reaching a value of 2.2 for a 246 nm plate. Te/Se ratio, which can be evaluated over the entire investigated thickness range, shows two distinct regimes: below 49 nm, it decreases steeply with thickness, while above 49 nm, the decrease continues but with a markedly reduced slope.

Notably, this same thickness range coincides with the transition observed in the Raman data: as discussed above, the position of the low-wavenumber component of the $E^2_g$ band, which may be associated with compositional disorder in the anion sublattice, decreases markedly up to 49 nm and remains comparatively stable at larger thicknesses.

The apparent crossover in compositional evolution raises questions regarding its possible origin. A plausible primary driver is the incongruent sublimation of the BSTS source. For the structurally related compound $Bi_2Te_3$, it was reported that sublimation of the bulk material proceeds incongruently, resulting in a higher partial pressure of $Te_2$ relative to BiTe and other Bi-rich vapor species [37]. Consequently, $Te_2$ has been identified as the first species to adsorb on a van der Waals substrate during the early stages of growth, specifically on $WS_2$, a layered semiconductor structurally analogous to $MoS_2$ [37]. By analogy, preferential release of Te- and Se-rich vapor species from the BSTS source during the initial stages of growth, followed by the progressive incorporation of Bi-containing species as the source composition evolves, could account for the near absence of Bi and the correspondingly elevated Te/Se ratio, both observed in the thinnest platelets. This scenario is further supported by the relative volatility of the binary compounds. Despite the high vapor pressure of elemental Se, $Bi_2Se_3$ is markedly less volatile than $Bi_2Te_3$ [47]. This may result in delayed release of the Bi- and Se-rich fraction of the source relative to the more volatile Sb- and Te-rich fraction. Differences in the vapor pressures of the constituent species have also been reported to produce compositional variations during the vapor deposition of crystalline Ge–Sb–Te films [48, 49]. Up to a platelet thickness of 49 nm, the deposition process appears to be strongly influenced by the balance between vapor-phase supply and re-evaporation from the growing surface, as reflected by the pronounced variation in elemental ratios with platelet thickness. Above this thickness, net incorporation into the crystalline phase appears to become more dominant, resulting in a more gradual evolution of both platelet thickness and composition. This behavior may indicate a transition from an initial growth regime, strongly affected by vapor–surface interactions and changing vapor composition, to a more stable growth regime in which the compositional evolution becomes less pronounced.

The fitted trends in Figure 5(c) indicate that the Bi/Sb and Te/Se ratios approach their nominal source values at different platelet thicknesses. The EDS measurements (Table S1) show that Bi/Sb values close to 1 are mainly observed for platelets with thicknesses of 120.5- 157 nm, whereas Te remains in excess relative to Se within this range. At larger thicknesses, the Te/Se ratio progressively approaches unity, while the Bi/Sb ratio increases above 1, indicating an increasing Bi excess relative to Sb.

The correlation between vibrational band positions and the Bi/Sb and Te/Se ratios shown in Figure 5(d) reveals the most significant variation again observed for the $A^2_{1g}$ mode. As discussed above, this mode appears to be the most sensitive to the local environment, arising from out-of-plane vibrations involving the pnictogen (Bi/Sb) and the external chalcogen (Te/Se) atoms. In addition, over the Te/Se ratio range from 1.03 to 2.8, a slight decrease in peak position is observed. Decreasing Se content in the platelets leads to a red shift of $A^2_{1g}$ mode. This is in good agreement with reports that substitution of Te by Se should induce a positive frequency shift, since Se atoms are lighter, smaller in size, and more electronegative than Te, resulting in shortening of the Se-Bi chemical bonds in $BiSb(Te_{1-y}Se_y)_3$ nanocrystals [22], as well as with another report showing that substitution of Se and S atoms at the Te site shifts the vibrational frequency to higher values in ternary tetradymites [29]. The low-wavenumber component of the $E^2_g$ band decreases from 104.7 cm$^{-1}$ at a Te/Se ratio of 9.2 to 97.7 cm$^{-1}$ at a ratio of 4.1 and thereafter remains relatively constant in position. This trend further supports its composition-sensitive nature.

**4. Conclusion**

Van der Waals BSTS platelets were successfully grown on a semiconducting $MoS_2$ substrate *via* a simple vapor-phase process conducted under reduced pressure and without toxic organometallic precursors. SEM analysis revealed an average lateral size of 1.4 μm, while detailed correlative characterization was performed on platelets with thicknesses ranging from 25 to 246 nm. These results demonstrate that $MoS_2$ supports the growth of BSTS platelets with micrometer-scale lateral dimensions and nanometer-scale thicknesses, thereby providing a technologically relevant semiconducting alternative to electrically insulating substrates.

Detailed structural and compositional characterization revealed pronounced disorder in the grown BSTS plates, consistent with the quaternary solid-solution nature of the material, as evidenced by both additional disorder-related Raman features and thickness-dependent compositional gradients. Raman spectroscopy performed across BSTS plates showed contributions from both BSTS and the underlying $MoS_2$ substrate, demonstrating that the substrate Raman response remains optically accessible even through relatively thick BSTS plates. Among the characteristic BSTS modes, $A^2_{1g}$ was found to be the most sensitive to both plate thickness and local composition. EDS analysis revealed the thickness-dependent variation in the Bi/Sb and Te/Se atomic ratios, most likely related to the different volatilities of the constituent elements released from the solid source during growth.

The present results provide a structural and compositional foundation for further investigation of BSTS/$MoS_2$ heterostructures. Future work should focus on optimizing the growth conditions, particularly the source composition, to reduce the observed compositional variations. In this context, systematically varying $x$ and $y$ would help clarify how source composition influences the growth and properties of BSTS platelets on $MoS_2$. Improved control over platelet thickness may be achieved through more uniform, large-area continuous $MoS_2$ substrates, better-controlled vapor transport through regulated carrier-gas flow, and optimization of the duration of the nucleation and growth stages. In addition, direct electrical transport and magnetotransport measurements of individual BSTS plates will be important for determining whether the topological surface states are preserved on the $MoS_2$ substrate and whether the BSTS/$MoS_2$ interface induces measurable charge transfer or band bending.

**References**

[1] F. Ortmann, S. Roche, S.O. Valenzuela, eds., Topological Insulators: Fundamentals and Perspectives, 1st ed., Wiley, 2015. https://doi.org/10.1002/9783527681594.

[2] Y. Zheng, T. Xu, X. Wang, Z. Sun, B. Han, Study on Bulk-Surface Transport Separation and Dielectric Polarization of Topological Insulator $Bi_{1.2}Sb_{0.8}Te_{0.4}Se_{2.6}$, Mol. 29 (2024) 859. https://doi.org/10.3390/molecules29040859.

[3] S. Pal, A. Nandi, S.G. Nath, P.K. Pal, K. Sharma, S. Manna, A. Barman, C. Mitra, Enhancement of spin to charge conversion efficiency at the topological surface state by inserting normal metal spacer layer in the topological insulator based heterostructure, Appl. Phys. Lett. 124 (2024) 112416. https://doi.org/10.1063/5.0192717.

[4] H.-S. Kim, H.S. Shin, J.S. Lee, C.W. Ahn, J.Y. Song, Y.-J. Doh, Quantum electrical transport properties of topological insulator $Bi_2Te_3$ nanowires, Curr. Appl Phys. 16 (2016) 51–56. https://doi.org/10.1016/j.cap.2015.10.011.

[5] O. Ivanov, M. Yaprintsev, A. Vasil'ev, E. Yaprintseva, Microstructure and thermoelectric properties of the medium-entropy block-textured $BiSbTe_{1.5}Se_{1.5}$ alloy, J. Alloys Compd. 872 (2021) 159743. https://doi.org/10.1016/j.jallcom.2021.159743.

[6] [H. Li, L. Sang, M. Li, G. Peng, J. Chen, N. Ye, G. Yang, X. Wang, Optimizing thermoelectric performance of medium-entropy $(Bi,Sb)_2(Te,Se)_3$ topological insulators via bulk band warping, J. Alloys Compd. 1042 (2025) 184010. https://doi.org/10.1016/j.jallcom.2025.184010.

[7] M.Y. Toriyama, G.J. Snyder, Topological insulators for thermoelectrics: A perspective from beneath the surface, The Innovation 6 (2025) 100782. https://doi.org/10.1016/j.xinn.2024.100782.

[8] A. Vasil'ev, M. Zhezhu, O. Ivanov, Photoresponse properties of single-crystalline thick film based on high-entropy topological insulator $(Bi_{3/4}Sb_{1/4})_2(Te_{2/5}Se_{2/5}S_{1/5})_3$, Mater. Lett. 402 (2026) 139293. https://doi.org/10.1016/j.matlet.2025.139293.

[9] Z. Liu, Z. Guo, A. Li, L. Wang, J. Sui, T. Mori, Tellurium-free thermoelectric materials and devices for low-temperature energy harvesting, Nat Rev Mater (2026). https://doi.org/10.1038/s41578-026-00923-5.

[10] Z. Guo, L. Wang, Z. Xu, Y.-K. Zhu, X. Dong, H. Wu, F. Guo, W. Cai, J. Sui, Z. Liu, Dopant-dependent pore formation in plastic Ag2Se contributing to ultrahigh thermoelectric performance, Acta Materialia 306 (2026) 121917. https://doi.org/10.1016/j.actamat.2026.121917.

[11] H. Lohani, P. Mishra, A. Banerjee, K. Majhi, R. Ganesan, U. Manju, D. Topwal, P.S.A. Kumar, B.R. Sekhar, Band Structure of Topological Insulator $BiSbTe_{1.25}Se_{1.75}$, Sci. Rep. 7 (2017) 4567. https://doi.org/10.1038/s41598-017-04985-y.

[12] W. Wang, L. Li, W. Zou, L. He, F. Song, R. Zhang, X. Wu, F. Zhang, Intrinsic Topological Insulator $Bi_{1.5}Sb_{0.5}Te_{3-x}Se_x$ Thin Crystals, Sci. Rep. 5 (2015) 7931. https://doi.org/10.1038/srep07931.

[13] K.-B. Han, S.K. Chong, A.O. Oliynyk, A. Nagaoka, S. Petryk, M.A. Scarpulla, V.V. Deshpande, T.D. Sparks, Enhancement in surface mobility and quantum transport of $Bi_{2-x}Sb_xTe_{3-y}Se_y$ topological insulator by controlling the crystal growth conditions, Sci. Rep. 8 (2018) 17290. https://doi.org/10.1038/s41598-018-35674-z.

[14] N. Kumar, D.V. Ishchenko, I.A. Milekhin, A.T. Kozakov, A.V. Nikolskii, S.V. Pryanichnikov, A.G. Milekhin, O.E. Tereshchenko, Temperature-dependent anomalous renormalization of phonon frequency and linewidth of MBE-grown $Bi_2Te_3$ and $Bi_{0.8}Sb_{1.2}Te_{1.8}Se_{1.2}$ topological insulator thin films investigated by Raman spectroscopy, Surf. Interfaces 95 (2026) 109776. https://doi.org/10.1016/j.surfin.2026.109776.

[15] Quantum transport study in three-dimensional topological insulator $BiSbTeSe_2$, in: Semiconductors and Semimetals, Elsevier, 2021: pp. 73–124. https://doi.org/10.1016/bs.semsem.2021.07.002.

[16] T. Xu, Y. Zheng, X. Wang, Z. Sun, B. Han, Study of dielectric polarization and electrical transport in $Bi_{1.2}Sb_{0.8}Te_{0.4}Se_{2.6}$ nanofilms, Heliyon 10 (2024) e27444. https://doi.org/10.1016/j.heliyon.2024.e27444.

[17] J.P. Heremans, R.J. Cava, N. Samarth, Tetradymites as thermoelectrics and topological insulators, Nat Rev. Mater. 2 (2017) 17049. https://doi.org/10.1038/natrevmats.2017.49.

[18] M. Brahlek, N. Koirala, N. Bansal, S. Oh, Transport properties of topological insulators: Band bending, bulk metal-to-insulator transition, and weak anti-localization, Solid State Commun. 215–216 (2015) 54–62. https://doi.org/10.1016/j.ssc.2014.10.021.

[19] N. Koirala, M. Brahlek, M. Salehi, L. Wu, J. Dai, J. Waugh, T. Nummy, M.-G. Han, J. Moon, Y. Zhu, D. Dessau, W. Wu, N.P. Armitage, S. Oh, Record Surface State Mobility and Quantum Hall Effect in Topological Insulator Thin Films via Interface Engineering, Nano Lett. 15 (2015) 8245–8249. https://doi.org/10.1021/acs.nanolett.5b03770.

[20] Ren, A.A. Taskin, S. Sasaki, K. Segawa, Y. Ando, Optimizing $Bi_{2-x}Sb_xTe_{3-y}Se_y$ solid solutions to approach the intrinsic topological insulator regime, Phys. Rev. B 84 (2011) 165311. https://doi.org/10.1103/PhysRevB.84.165311.

[21] N.H. Tu, Y. Tanabe, K.K. Huynh, Y. Sato, H. Oguro, S. Heguri, K. Tsuda, M. Terauchi, K. Watanabe, K. Tanigaki, Van der Waals epitaxial growth of topological insulator $Bi_{2-x}Sb_xTe_{3-y}Se_y$ ultrathin nanoplate on electrically insulating fluorophlogopite mica, Appl. Phys. Lett. 105 (2014) 063104. https://doi.org/10.1063/1.4892576.

[22] N. Abdelrahman, T. Charvin, S. Froeschke, R. Giraud, J. Dufouleur, A. Popov, S. Schiemenz, D. Wolf, B. Büchner, M. Mertig, S. Hampel, Controlled growth of 3D topological insulator $BiSb(Te_{1-y}Se_y)_3$ nanocrystals *via* chemical vapor transport, J. Mater. Chem. C 12 (2024) 18416–18426. https://doi.org/10.1039/D4TC02508C.

[23] S. Mukhopadhyay, P.K. Pal, S. Manna, C. Mitra, A. Barman, All-optical observation of giant spin transparency at the topological insulator $BiSbTe_{1.5}Se_{1.5}/Co_{20}Fe_{60}B_{20}$ interface, NPG Asia Mater. 15 (2023) 57. https://doi.org/10.1038/s41427-023-00504-w.

[24] N. Kumar, N.V. Surovtsev, P.A. Yunin, D.V. Ishchenko, I.A. Milekhin, S.P. Lebedev, A.A. Lebedev, O.E. Tereshchenko, Raman scattering spectroscopy of MBE grown thin film topological insulator $Bi_{2-x}Sb_xTe_{3-y}Se_y$, Phys. Chem. Chem. Phys. 26 (2024) 13497–13505. https://doi.org/10.1039/D4CP01169D.

[25] A.S. Tarasov, N. Kumar, V.A. Golyashov, I.O. Akhundov, D.V. Ishchenko, K.A. Kokh, A.O. Bazhenov, N.P. Stepina, O.E. Tereshchenko, Formation of well-ordered surfaces of $Bi_{2-x}Sb_xTe_{3-y}Se_y$ topological insulators using wet chemical treatment, Appl. Surf. Sci. 649 (2024) 159122. https://doi.org/10.1016/j.apsusc.2023.159122.

[26] P I Kuznetsov, G G Yakushcheva, B S Shchamkhalova, V A Jitov, A G Temiryazev, V E Sizov, V O Yapaskurt, MOVPE growth and transport characterization of $Bi_{2-x}Sb_xTe_{3-y}Se_y$ films, J. Cryst. Growth 483 (2018) 216-222. https://doi.org/10.13140/RG.2.2.10879.5176.

[27] A.R. Cadore, R. De Oliveira, R. Longuinhos, V.D.C. Teixeira, D.A. Nagaoka, V.T. Alvarenga, J. Ribeiro-Soares, K. Watanabe, T. Taniguchi, R.M. Paniago, A. Malachias, K. Krambrock, I.D. Barcelos, C.J.S. De Matos, Exploring the structural and optoelectronic properties of natural insulating phlogopite in van der Waals heterostructures, 2D Mater. 9 (2022) 035007. https://doi.org/10.1088/2053-1583/ac6cf4.

[28] K.H.M. Chen, H.Y. Lin, S.R. Yang, C.K. Cheng, X.Q. Zhang, C.M. Cheng, S.F. Lee, C.H. Hsu, Y.H. Lee, M. Hong, J. Kwo, Van der Waals epitaxy of topological insulator $Bi_2Se_3$ on single layer transition metal dichalcogenide $MoS_2$, Appl. Phys. Lett. 111 (2017) 083106. https://doi.org/10.1063/1.4989805.

[29] M. Kanagaraj, A. Pawbake, S.Ch. Sarma, V. Rajaji, C. Narayana, M.-A. Measson, S.C. Peter, Structural, magnetotransport and Hall coefficient studies in ternary $Bi_2Te_2Se$, $Sb_2Te_2Se$ and $Bi_2Te_2S$ tetradymite topological insulating compounds, J. Alloys Compd. 794 (2019) 195–202. https://doi.org/10.1016/j.jallcom.2019.04.226.

[30] R. German, E.V. Komleva, P. Stein, V.G. Mazurenko, Z. Wang, S.V. Streltsov, Y. Ando, P.H.M. Van Loosdrecht, Phonon mode calculations and Raman spectroscopy of the bulk-insulating topological insulator $BiSbTeSe_2$, Phys. Rev. Materials 3 (2019) 054204. https://doi.org/10.1103/PhysRevMaterials.3.054204.

[31] J. Zhang, Z. Peng, A. Soni, Y. Zhao, Y. Xiong, B. Peng, J. Wang, M.S. Dresselhaus, Q. Xiong, Raman Spectroscopy of Few-Quintuple Layer Topological Insulator $Bi_2Se_3$ Nanoplatelets, Nano Lett. 11 (2011) 2407–2414. https://doi.org/10.1021/nl200773n.

[32] N. Kumar, N.V. Surovtsev, D.V. Ishchenko, P.A. Yunin, I.A. Milekhin, O.E. Tereshchenko, A.G. Milekhin, Resonance Raman Scattering of Topological Insulators $Bi_2Te_3$ and $Bi_{2-x}Sb_xTe_{3-y}Se_y$ Thin Films, J Raman Spectrosc. 56 (2025) 207–217. https://doi.org/10.1002/jrs.6751.

[33] T. Livneh, J.E. Spanier, A comprehensive multiphonon spectral analysis in $MoS_2$, 2D Mater. 2 (2015) 035003. https://doi.org/10.1088/2053-1583/2/3/035003.

[34] P. Ni, M. Dieng, J.-C. Vanel, I. Florea, F.Z. Bouanis, A. Yassar, Liquid Shear Exfoliation of $MoS_2$: Preparation, Characterization, and $NO_2$-Sensing Properties, Nanomater. 13 (2023) 2502. https://doi.org/10.3390/nano13182502.

[35] Y. Jiang, X. Zhang, Y. Wang, N. Wang, D. West, S. Zhang, Z. Zhang, Vertical/Planar Growth and Surface Orientation of $Bi_2Te_3$ and $Bi_2Se_3$ Topological Insulator Nanoplates, Nano Lett. 15 (2015) 3147–3152. https://doi.org/10.1021/acs.nanolett.5b00240.

[36] P. Wu, H. Chen, C. Yang, W. Gan, Z. Muhammad, L. Song, Tuning the composition of ternary $Bi_2Se_{3x}Te_{3(1-x)}$ nanoplates and their Raman scattering investigations, AIP Adv. 6 (2016) 075303. https://doi.org/10.1063/1.4958712.

[37] E. Kahn, M. Lucking, F. Zhang, Y. Lei, T. Granzier-Nakajima, D. Grasseschi, K. Beach, W. Murray, Y.-T. Yeh, A.L. Elias, Z. Liu, H. Terrones, M. Terrones, Selective Synthesis of $Bi_2Te_3/WS_2$ Heterostructures with Strong Interlayer Coupling, ACS Appl. Mater. Interfaces (2020) acsami.0c03656. https://doi.org/10.1021/acsami.0c03656.

[38] M.M. Hammo, M.A.A. Mohamed, E.K. Moustafa, A. Popov, D. Wolf, B. Büchner, M. Mertig, S. Hampel, Tunable chemical vapor transport growth of 2D $Bi_2$ $Te_3$ /$MoS_2$ van der Waals heterostructures, Nanoscale 18 (2026) 16622–16629. https://doi.org/10.1039/D6NR00967K.

[39] T. Ginley, Y. Wang, S. Law, Topological Insulator Film Growth by Molecular Beam Epitaxy: A Review, Crystals 6 (2016) 154. https://doi.org/10.3390/cryst6110154.

[40] A. Sharma, T.D. Senguttuvan, V.N. Ojha, S. Husale, Novel synthesis of topological insulator based nanostructures (Bi2Te3) demonstrating high performance photodetection, Sci Rep 9 (2019) 3804. https://doi.org/10.1038/s41598-019-40394-z.

[41] K.A. Kokh, V.V. Atuchin, T.A. Gavrilova, N.V. Kuratieva, N.V. Pervukhina, N.V. Surovtsev, Microstructural and vibrational properties of PVT grown $Sb_2Te_3$ crystals, Solid State Commun. 177 (2014) 16–19. https://doi.org/10.1016/j.ssc.2013.09.016.

[42] J. Yuan, M. Zhao, W. Yu, Y. Lu, C. Chen, M. Xu, S. Li, K. Loh, B. Qiaoliang, Raman Spectroscopy of Two-Dimensional $Bi_2Te_xSe_{3-x}$ Platelets Produced by Solvothermal Method, Mater. 8 (2015) 5007–5017. https://doi.org/10.3390/ma8085007.

[43] D. Sharma, M.M. Sharma, R.S. Meena, V.P.S. Awana, Raman spectroscopy of $Bi_2Se_{3-x}Te_x$ (x = 0–3) topological insulator crystals, Physica B. 600 (2021) 412492. https://doi.org/10.1016/j.physb.2020.412492.

[44] X. Qi, W. Ma, X. Zhang, C. Zhang, Raman characterization and transport properties of morphology-dependent two-dimensional $Bi_2Te_3$ nanofilms, Appl. Surf. Sci. 457 (2018) 41–48. https://doi.org/10.1016/j.apsusc.2018.06.142.

[45] C. Wang, X. Zhu, L. Nilsson, J. Wen, G. Wang, X. Shan, Q. Zhang, S. Zhang, J. Jia, Q. Xue, In situ Raman spectroscopy of topological insulator $Bi_2Te_3$ films with varying thickness, Nano Res. 6 (2013) 688–692. https://doi.org/10.1007/s12274-013-0344-4.

[46] W. Richter, C.R. Becker, A Raman and far-infrared investigation of phonons in the rhombohedral $V_2$–$VI_3$ compounds $Bi_2Te_3$, $Bi_2Se_3$, $Sb_2Te_3$ and $Bi_2(Te_{1-x}Se_x)_3$ $(0 < x < 1)$, $(Bi_{1-y}Sb_y)_2Te_3$ $(0 < y < 1)$, Phys. Status Solidi B 84 (1977) 619–628. https://doi.org/10.1002/pssb.2220840226.

[47] H. Noro, K. Sato, H. Kagechika, The thermoelectric properties and crystallography of Bi-Sb-Te-Se thin films grown by ion beam sputtering, J. Appl. Phys. 73 (1993) 1252–1260. https://doi.org/10.1063/1.353266.

[48] M. Zhezhu, A. Vasil'ev, M. Yapryntsev, E. Ghalumyan, D.A. Ghazaryan, H. Gharagulyan, Chemical vapor deposition synthesis of $(GeTe)(Sb_2Te_3)$ gradient crystalline films as promising planar heterostructures, Mater. Sci. Semicond. Process. 199 (2025) 109863. https://doi.org/10.1016/j.mssp.2025.109863.

[49] M. Zhezhu, A. Vasil'ev, M. Sukiasyan, A. Kutuzyan, N. Anosov, M. Yaprintsev, D.A. Ghazaryan, H. Gharagulyan, Laser-induced optical reconfiguration in Ge–Sb–Te films with composition-dependent response, Mater. Sci. Eng. 333 (2026) 119731. https://doi.org/10.1016/j.mseb.2026.119731.

**Supplementary Information**

**Vapor-phase growth and characterization of van der Waals $Bi_{2-x}Sb_xTe_{3-y}Se_y$ platelets on semiconducting $MoS_2$**

M. Zhezhu[1*], V. Torosyan[1], V. Mehrabyan[1], A. Vasil'ev[1]

*[1]A.B. Nalbandyan Institute of Chemical Physics, Yerevan, Republic of Armenia*

**Correspondence should be addressed to marina.zhezhu@ichph.sci.am*

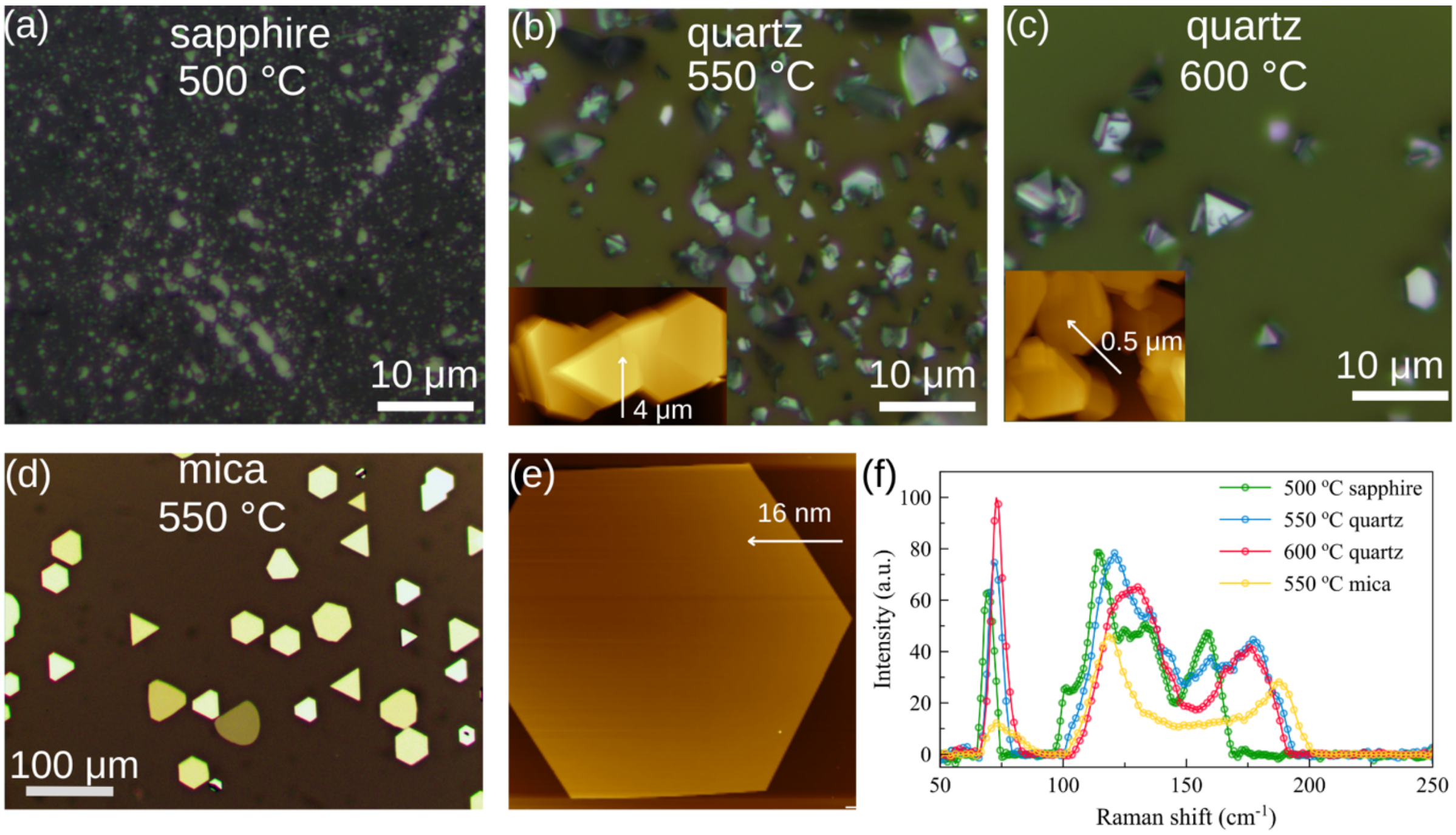

Figure S1. BSTS platelets obtained during optimization of the vapor phase deposition conditions: (a) growth at 500 °C on a sapphire substrate, (b) growth at 550 °C on a quartz substrate with the corresponding AFM image of typical platelets, (c) growth at 600 °C on a quartz substrate with the corresponding AFM image of typical platelets, and (d) growth at 550 °C on a mica substrate with (e) the corresponding AFM image of typical platelets. (f) Raman spectra of the obtained BSTS platelets.

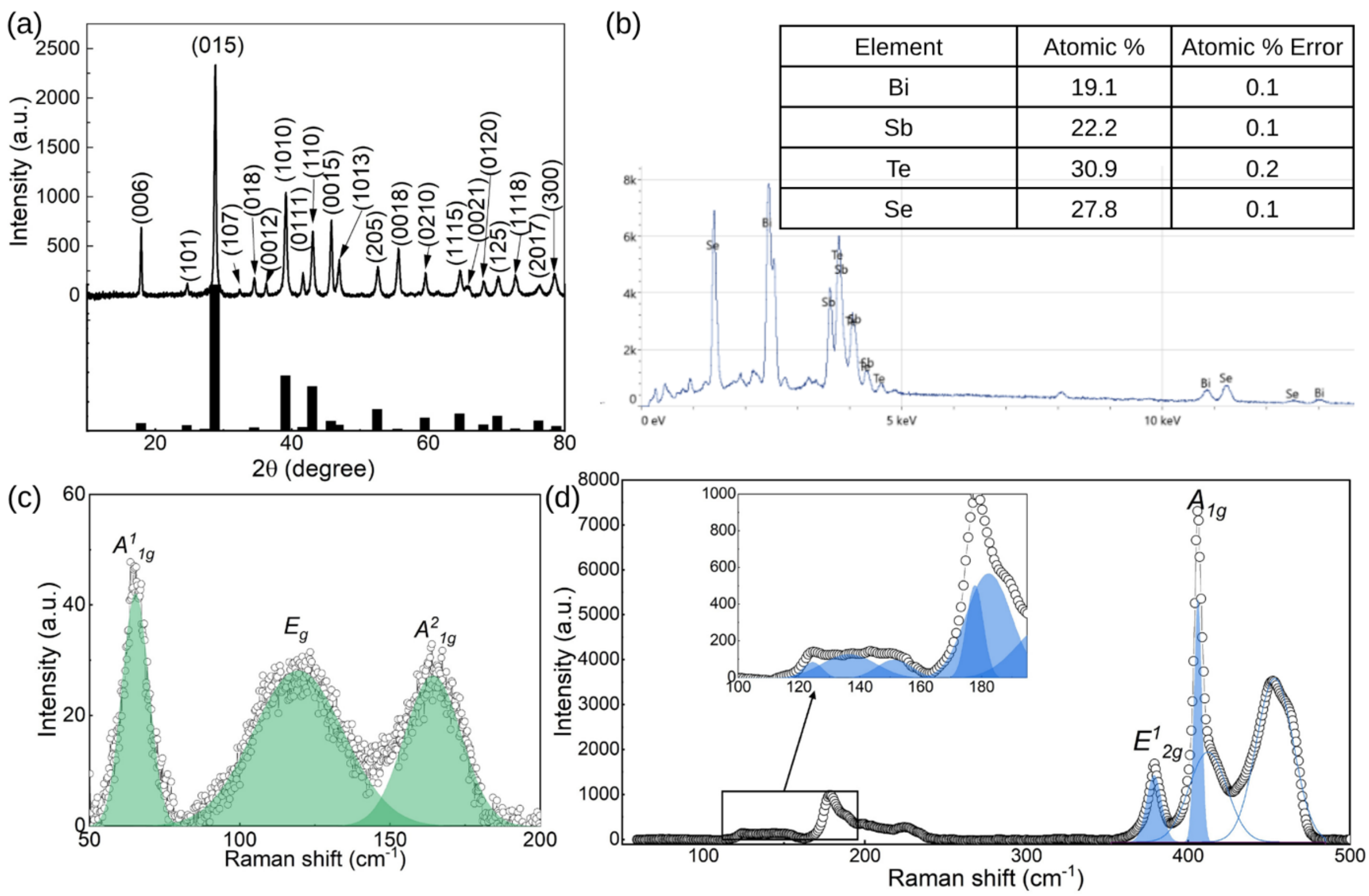

Figure S2. Characterization of the preliminary prepared and ground BSTS single-crystal source: (a) XRD pattern with the corresponding PDF-2 database reference card No. 01-082-5691, (b) EDS spectral profile and atomic composition, (c) Raman spectrum of BSTS with Gauss peak deconvolution highlighted in green. (d) Raman spectrum of the cleaved $MoS_2$ substrate with an enlarged view of the 100–200 cm$^{-1}$ spectral region and Gauss-deconvoluted peaks highlighted in blue.

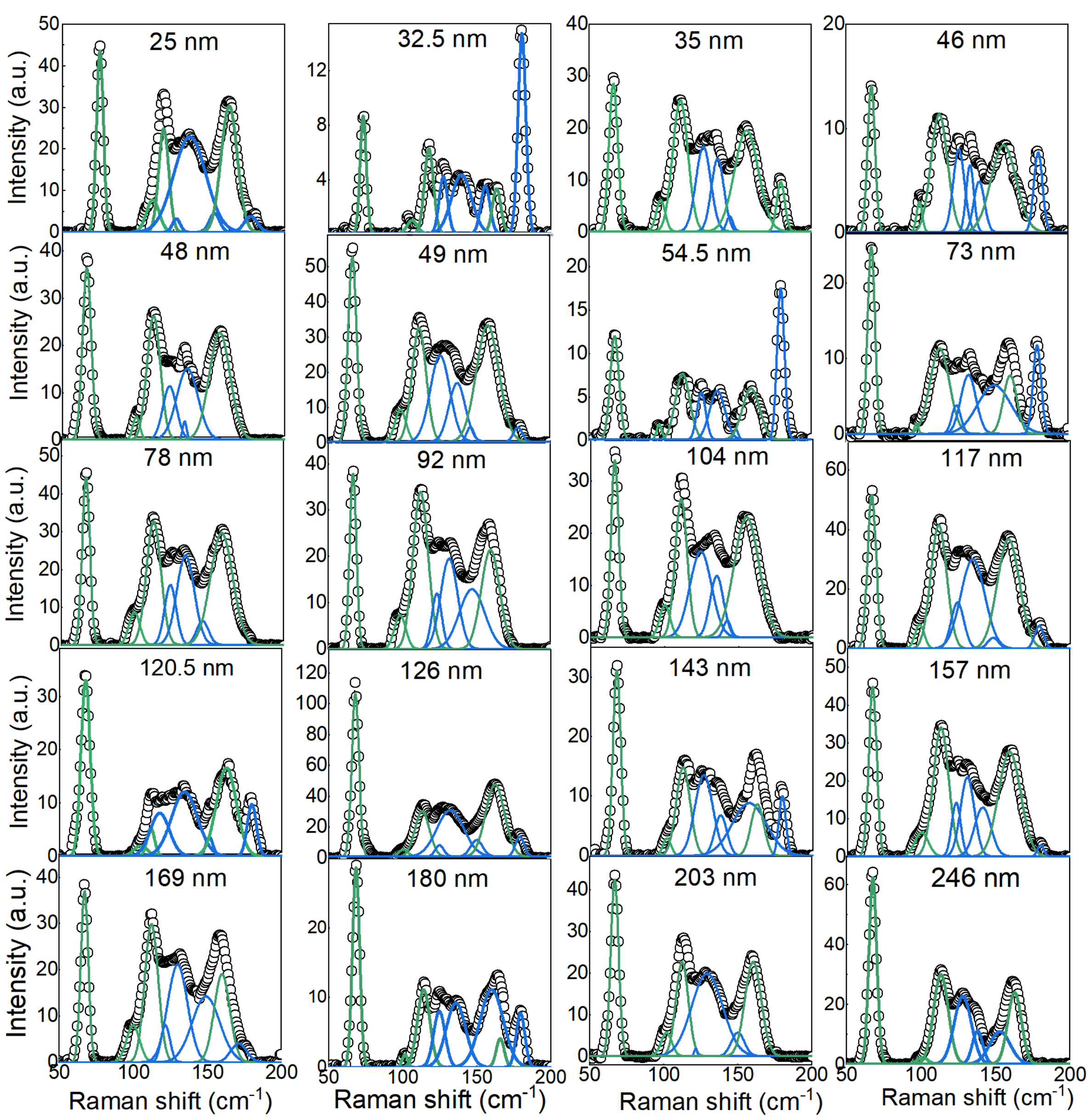

Figure S3. Gauss-deconvoluted Raman spectra of 20 van der Waals BSTS platelets with different thicknesses. Green peaks correspond to the characteristic BSTS vibrational modes and blue peaks to the vibrational modes of the $MoS_2$ substrate.

Table S1. Atomic composition of van der Waals epitaxial BSTS platelets determined by EDS analysis.

| Thickness, nm | 25 | 32.5 | 35 | 46 | 48 | 49 | 54.5 | 73 | 78 | 92 | 104 | 117 | 120.5 | 126 | 143 | 157 | 169 | 180 | 203 | 246 |
|---|---|---|---|---|---|---|---|---|---|---|---|---|---|---|---|---|---|---|---|---|
| Bi, at. % | n.q. | n.q. | n.q. | n.q. | n.q. | 0.1 ± 0.0 | 8.1 ± 0.3 | 6.3 ± 0.1 | 6.8 ± 0.3 | 9.3 ± 0.3 | 11.8 ± 0.2 | 11.9 ± 0.2 | 17.7 ± 0.2 | 12.2 ± 0.2 | 17.2 ± 0.2 | 14.8 ± 0.2 | 20.2 ±0.3 | 19.8 ±0.3 | 22.1 ±0.3 | 24.4 ±0.3 |
| Sb, at. % | 25± 0.3 | 27.8 ± 0.5 | 27.6 ± 0.6 | 24.9 ± 0.5 | 26 ± 0.5 | 24.9 ± 0.5 | 19.3 ± 0.5 | 24.6 ± 0.3 | 21.2 ± 0.5 | 18.7 ± 0.5 | 18 ± 0.5 | 18.5 ± 0.4 | 14.9 ± 0.4 | 16.9 ± 0.4 | 16.2 ± 0.2 | 16.4 ± 0.4 | 12.3 ±0.4 | 14.6 ±0.4 | 11.4 ±0.4 | 11.3 ±0.3 |
| Te, at. % | 62.4 ± 0.3 | 65.1 ± 0.6 | 63.9 ± 0.6 | 61.7 ± 0.6 | 62 ± 0.6 | 59.9 ± 0.6 | 49.9 ± 0.5 | 55.6 ± 0.3 | 53.1 ± 0.5 | 48.7 ± 0.5 | 47.8 ± 0.5 | 46.1 ± 0.5 | 40.7 ± 0.4 | 43.6 ± 0.4 | 43.7 ± 0.2 | 42.6 ± 0.4 | 36.5 ±0.4 | 37.6 ±0.4 | 34.1 ±0.4 | 32.7 ±0.4 |
| Se, at. % | 12.6 ±0.1 | 7.1± 0.3 | 8.5 ± 0.2 | 13.5 ± 0.3 | 12 ± 0.3 | 15.1 ± 0.3 | 22.7 ± 0.3 | 13.5 ± 0.1 | 18.9 ± 0.3 | 23.3 ± 0.3 | 22.4 ± 0.3 | 23.5 ± 0.2 | 26.6 ± 0.3 | 27.2 ± 0.3 | 22.9 ± 0.1 | 26.2 ± 0.3 | 31 ± 0.3 | 28 ± 0.3 | 32.4 ±0.3 | 31.7 ±0.3 |

*n.q. - not reliably quantifiable under the present EDS measurement conditions